\documentclass[aps,physrev,groupedaddress,twocolumn]{revtex4-2}

\usepackage{makecell}
\usepackage[usenames,dvipsnames]{pstricks}
\usepackage{floatrow}
\usepackage{mathtools}
\usepackage{enumitem}

\usepackage{color}
\usepackage{hyperref}

\usepackage{amsmath,amssymb,amsfonts,dsfont}

\usepackage[english]{babel}
\usepackage[utf8]{inputenc}
\usepackage{graphicx}

\usepackage{subfigure}
\usepackage{physics}
\usepackage{etex}

\usepackage{xcolor}
\usepackage{pgfplots}

\usepackage{amsthm}
\usepackage{tikz}
\usetikzlibrary{quantikz}
\usetikzlibrary{decorations.text, arrows.meta,calc,shadows.blur,shadings,intersections}

\newtheorem{theorem}{Theorem}
\newtheorem{definition}{Definition}

\definecolor{green}{RGB}{0, 128, 0}

\renewcommand{\L}{\mathcal{L}}
\newcommand{\M}{\mathcal{M}}

\newcommand{\infinity}{\infty}

\newcommand{\req}{\geq}

\newcommand{\term}[4]{{#1_{#2},#1_{#3},...,#1_{#4}}}
\newcommand{\Jlabel}{\mathbf{j}}
\newcommand{\Jcomp}{j}
\newcommand{\JcompX}[1]{ \Jcomp_{#1}}
\newcommand{\Jcompx}{ \JcompX{x}}

\newcommand{\Jcompindex}{\Jcompx=\term{k}{1}{2}{N-i}}	%vec notation

\newcommand{\sigi}{\sigma_i}
\newcommand{\MM}{N-2}  %Number of particles that can be lost (N-2 (N-2/2) and N in the distinguishable and undistinguishable distribution of W and GHZ (graph) states )
\newcommand{\LB}{\expval{\hat{E}}_{r,\psi}(\epsilon)}
\newcommand{\LBW}{\expval{\hat{E}}_{W}}

\newcommand{\expErpsi}{\expval{E}_{\mathbf{r},\psi}} %
\newcommand{\EIxprime}[1]{\expval{E}_{{#1}}' } %
\newcommand{\EIxprimeW}{\expval{\hat{E}}_{{W}}' } %

\newcommand{\EIpsi}{\expval{E}_{\psi}} %
\newcommand{\EIphi}{\expval{E}_{\phi}} %
\newcommand{\EIw}{\expval{\hat{E}}_{W}} %
\newcommand{\EIghz}{\expval{E}_{\ghz}} %
\newcommand{\EIG}{\expval{E}_{G}} %

\newcommand{\barE}[2]{\bar{E}^{*}_{{#1},{#2}} }
\newcommand{\barEum}[1]{\bar{E}^{*}_{#1} }

\newcommand{\dexpE}[1]{ \dv{\expval{E}_{{#1}}(\epsilon)}{\epsilon}}				
\newcommand{\dotE}[1]{ \expval{E}_{{#1}}'(\epsilon)}				
\newcommand{\dotEW}{ \expval{\hat{E}}_{{W}}'(\epsilon)}
\newcommand{\ghz}{\text{GHZ}}

\newcommand{\param}[1]{\vb*{\kappa}_{#1}}
\newcommand{\Mk}[1]{M_{k_{#1}}^{\param{#1}}  }

\newcommand{\wt}[1]{\abs{#1}}

\begin{document}

%\preprint{}

\title{Entanglement Distribution in Lossy Quantum Networks}


\author{Leonardo Oleynik}
\email[]{leonardo.oleynik@uni.lu}
\affiliation{Interdisciplinary Centre for Security, Reliability and Trust (SnT) – University of Luxembourg, L-1855 Luxembourg}

\author{Junaid ur Rehman}
\email[]{junaid.urrehman@kfupm.edu.sa}
\affiliation{Department of Electrical Engineering, and the Center for Intelligent Secure Systems, King Fahd University of Petroleum and Minerals (KFUPM), Dhahran 31261, Saudi Arabia}

\author{Seid Koudia}
\email[]{seid.koudia@uni.lu}
\affiliation{Interdisciplinary Centre for Security, Reliability and Trust (SnT) – University of Luxembourg, L-1855 Luxembourg}

\author{Symeon Chatzinotas}
\email[]{symeon.chatzinotas@uni.lu}
\affiliation{Interdisciplinary Centre for Security, Reliability and Trust (SnT) – University of Luxembourg, L-1855 Luxembourg}
% \IEEEmembership{Fellow, IEEE}


\date{\today}

\begin{abstract}
Entanglement distribution is essential for unlocking the potential of distributed quantum information processing. We consider an $N$-partite network where entanglement is distributed via a central source over lossy channels, and network participants cooperate to establish entanglement between any two chosen parties under local operations and classical communication (LOCC) constraints. We develop a general mathematical framework to assess the optimal average bipartite entanglement shared in a lossy distribution, and introduce a tractable lower bound by optimizing over a subset of single-parameter LOCC transformations. Our results show that probabilistically extracting Bell pairs from W states is more advantageous than deterministically extracting them from GHZ-like states in lossy networks, with this advantage increasing with network size. We further extend our analysis analytically, proving that W states remain more effective in large-scale networks. These findings offer valuable insights into the practical deployment of near-term networks, revealing a fundamental trade-off between deterministic entanglement distribution protocols and loss-sensitive resources.
\end{abstract}

%\keywords{}

\maketitle

\section{\label{sec:intro}Introduction}
	
	Quantum entanglement is fundamental for realizing the potential of distributed quantum information processing (DQIP). In this context, entanglement can be pictured as a resource that can be harnessed by physically separated parties, constrained by local operations and classical communication (LOCC), to perform various informational tasks. In such LOCC transformations, each party is allowed to measure their part of the system and broadcast the outcomes through a classical channel. This broadcasted information can subsequently inform updates in the measurements of other parties. Understanding the capabilities and limitations of such operations is crucial, as many quantum information tasks, including teleportation \cite{tele93}, one-way quantum computation \cite{OWQC01}, quantum conference key agreement \cite{21ConfeKey,23ConfeKeyNature}, and entanglement distribution \cite{12distGraph,19distGraph,SeidStabilizer}, rely on the LOCC paradigm.

	One-shot \emph{Random}-party entanglement distillation (RED) is a critical problem in DQIP. It involves investigating protocols to transform a single multipartite entangled state into a Bell pair shared among \emph{unspecified} parties. Contrary to multiple-copies entanglement distillation \cite{24distCap,21distStabilizer,10distconv}, is not possible to do quantum purification in one-shot settings. The advantages of random-party over specified-party entanglement distillation protocols were first highlighted in \cite{07FLWprotocol,08HKlo,11HK_lo}. In particular, it was shown that W states could be reliably converted to Bell pairs in a random-party entanglement distillation protocol, with a probability of success asymptotically reaching the unity for infinitely many rounds of LOCC operations \cite{07FLWprotocol}. However, subsequent findings demonstrated that LOCC operations cannot achieve this limit \cite{12HKlo}, as they belong to a class of quantum operations that is not topologically closed \cite{17LOCC,11LOCC_I}. Moreover, only recent studies have investigated the LOCC round complexity of single-copy random-party entanglement distillation protocols \cite{23round}. These results highlight some subtleties and open problems of RED protocols, especially concerning LOCC round complexity in random-party entanglement distribution from W states.

	W and Greenberger-Horne-Zeilinger (GHZ) states represent two distinct, nonequivalent entanglement classes for three-qubit systems \cite{00WandGHZ,01WNrobust}. As claimed in \cite{23QRepeaters}, their distinction lies in their entanglement structure: Although any bipartition of a GHZ state has maximal entanglement (exactly one ebit), the bipartite entanglement content of a W state is strictly less than one ebit. This structure results in two complementary aspects: while GHZ states can be \emph{deterministically} converted into Bell pairs (e.g., with a single qubit measurement in the Pauli-$X$ basis) but are sensitive to loss (all the entanglement is gone if any of the qubits are lost); W states can only be \emph{probabilistically} converted into Bell pairs, yet are robust to loss (some entanglement can be retrieved even if any of the qubits are lost). In this sense, for three-qubit systems, there is a clear tradeoff relationship between the success probability of entanglement conversion protocols for a given resource and the resource's robustness to losses. For instances, a deterministic conversion comes at the expense of loss-sensitivity, making each class preferable depending on the distributed information-processing scenario.

	Although entanglement distillation has been widely studied in both random and specified scenarios, few works have addressed these aspects in the context of near-term quantum networks \cite{25GHZinverse,21ConfeKey,23ConfeKeyNature}. Motivated by the tradeoff relationship between the success probability of entanglement conversion protocols for a given resource and the resource's robustness to losses present in three-partite entangled states, we developed a theoretical framework to assess bipartite entanglement conversion of arbitrary resources in a lossy network and employed it to compare W and GHZ-like states' performance in different network settings. This work examines the performance of single-copy, RED protocols in an $N$-partite centralized lossy network. Our primary contributions include: 
	\begin{itemize} 
		\item Developing a comprehensive theoretical framework to compute the average bipartite entanglement shared among a pair of parties within an $N$-partite centralized lossy network;
		
		\item Assessing single-parameter LOCC entanglement conversion performance for W and GHZ-like states by developing computationally efficient lower bounds for the figure of merit;
		
		\item Demonstrating W state's advantage over GHZ-like states numerically (and analytically for particular cases), corroborating the extension of a tradeoff relationship between the success probability of entanglement conversion protocols for a given resource and the resource's robustness to losses to multi-party states.
		
	\end{itemize} 
	The remainder of this paper is organized as follows. In Section~\ref{sec:PS}, we define the general problem, mathematical model, and notation, as well as our figure of merit and benchmark. In Section~\ref{sec:meth}, we introduce a tractable lower bound for our figure of merit and its simplifications for single-round, single-parameter (SP) LOCC transformations, applying it in Section~\ref{sec:app} to compare the performance of W and GHZ-like states in lossy entanglement distribution. We conclude our discussion and provide possible future directions in Section~\ref{sec:conc}.

	\section{\label{sec:PS}Problem statement: entanglement distribution in a lossy network}

	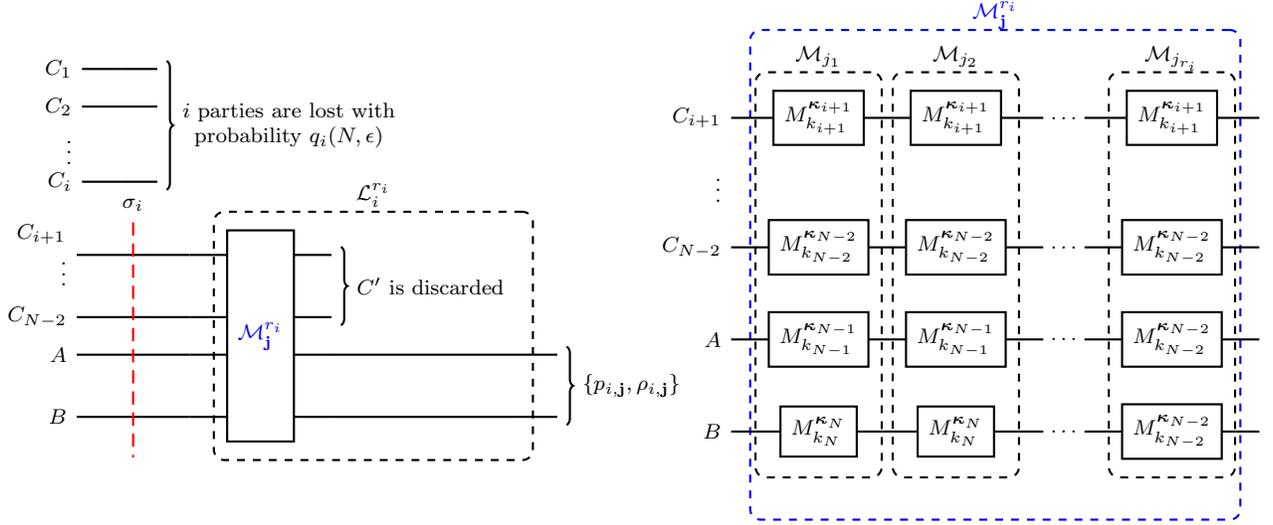
\begin{figure*}%[h]
		\centering
		\begin{tikzpicture}
%			\node (A) {	\input{ 0910_sumv3.tex} };
			\node (B)  {\begin{tikzpicture}[
	font=\footnotesize, 
	text centered,
	gategroup/.style ={dashed,rounded corners, inner xsep=2pt}
	]
	\def\x{1cm}\def\xx{.0cm}\def\y{3.5cm}
	\node   (A)                     {
		\begin{quantikz}
			\lstick{$C_1$}&[\xx]\qw& \qw\rstick[4]{$i$ parties are lost with\\ probability $q_i(N,\epsilon)$}\\
			\lstick{$C_2$}&[\xx]\qw& \qw\\ 
			\lstick{$\vdots$} &\\
			\lstick{$C_i$} &[\xx]\qw&\qw
		\end{quantikz}
	};
	\node   (B) [below= \y of A.south west, anchor=south west, xshift=-.5cm] {
		\begin{quantikz}
			\lstick{$C_{i+1}$\\$\quad\quad\vdots$} \slice{$\sigma_i$}&[\x]\qw& \gate[4]{\blue{\M_\Jlabel^{r_i}}} \gategroup[4,steps=3,style=gategroup]{$\L_i^{r_i}$} &\qw\rstick[2]{$C'$ is discarded} \\
			\lstick{$C_{N-2}$}&[\x]\qw&					&\qw\\
			\lstick{$A$}&[\x]\qw			&						&\qw&[2cm]\qw&\qw\rstick[2]{$\{ p_{i,\Jlabel}, \rho_{i,\Jlabel}  \}$}\\
			\lstick{$B$}&[\x]\qw			&						&\qw&[2cm]\qw&\qw
		\end{quantikz}
	};
\end{tikzpicture} } ;
			\node [right=-1cm of B] {\begin{tikzpicture}[
	font=\footnotesize, 
	text centered,
	gategroup/.style ={dashed,rounded corners, inner xsep=2pt},
	gategroupb/.style ={blue,dashed,rounded corners, inner ysep=20pt,inner xsep=4pt} 
	]
	\node {
		\begin{quantikz}
			\lstick{$C_{i+1}$}&\gate{\Mk{i+1}} \gategroup[5,steps=1,style=gategroup]{$\M_{j_1}$} 					\gategroup[5,steps=4,style=gategroupb,background]{\blue{$\M_\Jlabel^{r_i}$}} %erase 
			&\gate{\Mk{i+1}} \gategroup[5,steps=1,style=gategroup]{$\M_{j_2}$}  &\qw \ \ldots\ &\gate{\Mk{i+1}} \gategroup[5,steps=1,style=gategroup]{$\M_{j_{r_i}}$}  &\qw\\   %
			\lstick{$\vdots$} &\\
			\lstick{$C_{N-2}$}&\gate{\Mk{N-2}}&	\gate{\Mk{N-2}}							&\qw \ \ldots\	&\gate{\Mk{N-2}}&\qw\\
			\lstick{$A$}&\gate{\Mk{N-1}}&\gate{\Mk{N-1}}									&\qw \ \ldots\ &\gate{\Mk{N-2}}&\qw\\
			\lstick{$B$}&\gate{\Mk{N}}&\gate{\Mk{N}}										&\qw \ \ldots\ &\gate{\Mk{N-2}}&\qw
		\end{quantikz}
	};
\end{tikzpicture}

%\gategroup[2,steps=3,style={dashed,rounded
%	corners,fill=blue!20, inner
%	xsep=2pt},
%	background,
%	label style={label
%	position=below,anchor=north,yshift=-0.2cm}]

%\gategroup[2,steps=3,style={dashed,rounded
%	corners,fill=blue!20, inner
%	xsep=2pt}
 };
		\end{tikzpicture}
		\caption{\label{fig:sum}On the left, entanglement distribution and LOCC processing in a lossy network. On the the right, the sequence of $r_i$ global measurements $\M_{j_x}$, namely $\M_\Jlabel^{r_i}$, is detailed.}		
	\end{figure*}

	We consider an $N$-partite network that is served by a centralized source that generates and then distributes an $N$-qubit state (resource) $\psi$ to the network participants. The links from the source to each node are assumed to be lossy, i.e., a particle generated at the source gets lost in the link with probability $\epsilon$. Losses in all links are considered to be independent and identically distributed. The participants cooperate by measuring the received particles to generate entanglement between any two participants $A$ and $B$. Each node can perform $r_i\geq 1$ rounds of measurement on its received particle and broadcast the outcomes of each round to the network via a classical broadcast channel, i.e., the parties are constrained by LOCCs. 
	Given this entanglement distribution scenario, we want to:
	\begin{enumerate}
		\item Evaluate the amount of entanglement  shared on average between two network participants as a function of $\epsilon$, for a given parameter region $(N,r_i)$ and a resource state $\psi$;%, where $\mathbf{r}=(r_1,r_2,...,r_N)$; 
		
		\item Compare W and GHZ-like states' performance over different parameter regions, including asymptotic regimes (e.g., $N\to \infinity$). 
	\end{enumerate}
	Next, we formalize the above concepts and goals.

	\subsection{\label{subsec:model}Mathematical model and notations}

	We arbitrarily label the participants of interest by $A$ and $B$ and the helper participants by $C$ with an index, i.e., $C_1,C_2,...,C_i,...,C_{\MM}$ as depicted in Fig.~\ref{fig:sum}. Labeling is arbitrary because in a RED protocol success is deemed whenever some entangled bipartite state (not necessarily the maximally entangled one) is obtained between \emph{any} two parties \cite{23round,08HKlo,11LOCC_I}. In the distribution phase, any of the $N$ particles can be lost, but if more than $N-2$ are lost, nothing can be done locally to increase the entanglement. Then, the probability that $i\in \left\{0, 1, \cdots, \MM \right\}$ particles are lost is
	\begin{equation}
		q_i\left(N, \epsilon \right)=\binom{\MM}{i} \epsilon^i (1-\epsilon)^{\MM-i}.
		\label{eq:q_i}
	\end{equation}
	When exactly $i$ particles are lost, the overall state of the system is 
	\begin{equation}
		\sigma_i^N:= \Tr_{C_1, C_2,...C_i} \dyad{\psi} 
		\label{def:sig}
	\end{equation}
	(or simply $\sigma_i$ when the $\psi$'s dimension can be inferred from the context), where $C' $ is the partition comprising the remaining helpers $C_{i+1}, C_{i+2},..., C_{\MM}$.\footnote{We are considering resource states $\psi$ that are symmetric. Therefore, w.l.o.g., we can assume that the particles belonging to $C_1, \cdots C_i$ are lost during the distribution.} The received states and their corresponding probabilities form an ensemble 
	\begin{equation}
		\psi \to \{ q_i, \sigma_i \}.
		\label{eq:received}
	\end{equation}
	
	After the distribution phase, all parties, $C'$, $A$, and $B$, take turns measuring their local systems and broadcasting the results. Since each party holds a qubit, local measurements can be described by the $2\times 2$ Kraus operators $\{ \Mk{ }\}_k$. We can assume, without loss of generality \cite{23round}, that each $\Mk{}$ is in the upper triangular form
	\begin{equation}
		\Mk{} = 
		\begin{pmatrix}
			\sqrt{a_k} & b_k \\
			0 & \sqrt{c_k}
		\end{pmatrix},
		\label{eq:Kraus}
	\end{equation}
	where $\vb*{\kappa}=(a_k,b_k,c_k)$, $a_k,c_k\leq 0$, and the completion relation implies that $\sum_k a_k =1$ and $\sum_k c_k \req1$.

	Each received state $\sigma_i$ is processed by an \textit{$r_i$-round LOCC transformation} $\L_i^{r_i}$ that converts it into a state $\rho_{i,\Jlabel}$ with probability $p_{i,\Jlabel}$. In other words, an \textit{$r_i$-round LOCC transformation} can be viewed as a map
	\begin{equation}
		\L_i^{r_i}: \sigma_i \to \{ p_{i,\Jlabel}, \rho_{i,\Jlabel}  \},
		\label{eq:map}
	\end{equation}
	where
	\begin{align}
		p_{i,\Jlabel} &= \Tr      \mathcal{M}_\Jlabel^{r_i}      \mathcal{M}_\Jlabel^{r_i}       \sigi,  	\label{eq:rho}\\
		\rho_{i,\Jlabel} &= \frac{   \Tr_{C'} \mathcal{M}_\Jlabel^{r_i}  \sigi   \mathcal{M}_\Jlabel^{r_i}       }{p_{i,\Jlabel}}.
		\label{eq:map_exp}
	\end{align}
	In the above expression, $ \mathcal{M}_\Jlabel^{r_i}$ represents a sequence of $r_i$ global measurements given by the following composition rule
	\begin{equation}
		\Bqty{ \mathcal{M}_\Jlabel^{r_i} = \bigcirc_{x=1}^{r_i}   \M_{j_x}            }_{\Jlabel=j_{r_i},j_{r_i-1},...,j_1}
		\label{eq:rounds}
	\end{equation}
	wherein every $ \M_{\Jcompx} $ is expressed as the tensor product of LOCCs over the remaining $N-i$ subspaces, i.e.,
	\begin{equation}
		\Bqty{\M_{\Jcompx}  =    \otimes_{m=1}^{N-i} \Mk{m} }_{\Jcompindex}.
		\label{eq:globalMk}		
	\end{equation}
	in which $\Mk{m}$ has the form of \eqref{eq:Kraus}, parameterized by $\vb*{\kappa}_m=(a_{k_m},b_{k_m},c_{k_m})$. It is worth noting that, in general, the LOCCs performed in each node differ among the nodes $\vb*{\kappa}_m \neq \vb*{\kappa}_{m'}$ and from round to round $\M_{\Jcompx} \neq \M_{\JcompX{x'}}$. Likewise, each node can perform its LOCCs a different number of times $r_i \neq r_{i'}$. Later, in order to simplify our problem, we will consider a particular class of transformations where the LOCCs are identical in every round, identical for every system and each node is constrained by the same amount of rounds.

	We are interested in quantifying the amount of bipartite entanglement shared on average among $A$ and $B$ after $r_i$ rounds of LOCC transformations, i.e., the entanglement content of the reduced state $\rho_{i,\Jlabel}$. Since system labeling is arbitrary for RED protocols, we can always choose $AB $ as the bipartition with the maximum amount of entanglement and $C'$ as the remaining partitions. This implies in \emph{equivalent states}, i.e., reduced states with the same amount of bipartite entanglement. Next, we present the figure of merit that takes this into account as well as its benchmark.
	
	\subsection{Figure of merit and benchmark}
	
	We propose \emph{the average bipartite entanglement optimally shared among the target parties through a lossy network in $\mathbf{r}= (r_0, r_1,..., r_i,...,r_{N-2})$ rounds} as the figure of merit, defined as follows
	\begin{equation}
		\expErpsi(\epsilon) := \sum_i q_i(N,\epsilon) \barE{r_i}{\sigi},
		\label{eq:fig_psi}
	\end{equation}
	where
	\begin{equation}
		\barE{r_i}{\sigi}:= \sup_{   \L_i^{r_i} } \sum_{j} p_{i,\Jlabel} E(\rho_{i,\Jlabel}),
		\label{eq:fig_sig} 
	\end{equation}
	and $E$ is some bipartite entanglement measure, e.g., the entanglement of formation or concurrence \cite{01EoFandC}. We will adopt the latter in our numerical evaluations. 
		
	The definitions above are averages of the shared bipartite entanglement over the lossy distribution \eqref{eq:received} and the probabilisitic conversion \eqref{eq:map} distributions. More specifically, whereas \eqref{eq:fig_sig} is the average bipartite entanglement optimally achieved given a received state $\sigi$ via $\L_i^{r_i}$ protocols; \eqref{eq:fig_psi} is the average of \eqref{eq:fig_sig} over the ensemble of received states. Therefore, our definitions fully consider the statistical nature of the problem.

	A similar definition to \eqref{eq:fig_sig} is presented in \cite{23round} for a different entanglement distribution context. Applied to our distribution scenario, this definition could be written as $\sup_{r_i} \barE{r_i}{\sigi}$ and interpreted as the optimal average bipartite entanglement given any LOCC protocol, which includes protocols with an unbounded number of rounds. We adopted \eqref{eq:fig_sig} to avoid unbounded LOCC protocols and to distinguish the effect of the number of rounds in the figure of merit.

	According to \cite{12FLmono}, for finite-copy entanglement distillation, an entanglement \emph{measure} is any nonnegative function $E$ which is monotonically nonincreasing under LOCC transformations. Despite the lack of consensus on the necessary conditions that an entanglement measure must satisfy, monotonicity is often claimed as the essential property \cite{12FLmono}. In a quick inspection, we observe that \eqref{eq:fig_sig} is an entanglement measure by definition, while \eqref{eq:fig_psi} is also an entanglement measure as it represents the average of \eqref{eq:fig_sig}.
	
	We benchmark W states against two-centered GHZ graph states. As detailed in the Appendix~\ref{sec:ap_graph}, two-centered GHZ graph states have the property of being loss-robust. More precisely, it has been shown \cite{23robgGraph} that if multiple qubits adjacent to the same root are lost, the remaining state is always a GHZ state, and if qubits adjacent to both roots or the roots themselves are lost, the remaining state is fully separable. Such an entanglement structure simplifies our analysis, as after the distribution phase the received state is either a GHZ state, which can be deterministically converted to Bell pairs in a single round with a single measurement, or is fully separable, having no distillable entanglement whatsoever. In both cases, \eqref{eq:fig_sig} is trivially and exactly computed --- $\barE{r_0}{\sigma_0} = 1$ if none of the qubits is lost and $\barE{r_i}{\sigi} = 0$ if any qubit is lost.

	\section{\label{sec:meth}Methodology}

	In this section, we establish a tractable lower bound for the figure of merit by optimizing \eqref{eq:fig_psi} over the subset of single-parameter LOCC (SP-LOCC) transformations. We then analyze the properties of single-round, SP-LOCC transformations to derive explicit expressions for \eqref{eq:map_exp} and enhance computational efficiency in evaluating the lower bound.

	\subsection{Lower Bound}
	
	In the general formulation of the problem, the local operations performed are different from one node to the other and have the form \eqref{eq:Kraus} whose parameters $a_k, b_k, c_k$ can change in every round of the protocol, as emphasized in \ref{subsec:model}. For simplicity, we will restrict ourselves to the family of \emph{SP-LOCC transformations} $\L_i^{r,\kappa}$, where the measurements are
	\begin{enumerate}
		\item identical for every system, i.e., $\vb*{\kappa}_m = \vb*{\kappa}_{m'} \,\forall\, m,m' \in [1,N-i]$% 
		\item identical in every round, i.e., $\M_{\Jcompx} = \M_{j_{x'}} \,\forall\,  x$ and $x' \in [0,r_i]$   
		\item single parameterized $\vb*{\kappa}_m =\kappa_m $, i.e., the Kraus operators have the simplified form
		\begin{equation}
			M_0^\kappa=\begin{pmatrix}
				\sqrt{1-\kappa} & 0 \\
				0 & 1
			\end{pmatrix}, \quad\quad
			M_1^\kappa=\begin{pmatrix}
				\sqrt{\kappa} & 0 \\
				0 & 0
			\end{pmatrix},
			\label{eq:WKraus}
		\end{equation}
		where $\kappa \in [0,1]$ as proposed in \cite{07FLWprotocol}.
	\end{enumerate}
	Moreover, we assume the same number of rounds for every received state, i.e., $r_i=r \, \forall \, i$. These assumptions allow us to simplify the set of Kraus operators \eqref{eq:rounds} and \eqref{eq:globalMk} as follows
	\begin{equation}
		\Bqty{ \mathcal{M}_\Jlabel^r = \bigcirc_{x=1}^{r} \mathcal{M}_{j_x}
		}_{\Jlabel=j_r,j_{r-1},...,j_1}
		\label{eq:Mjlabel}
	\end{equation}
	with
	\begin{equation}
		\Bqty{ \mathcal{M}_{j_x} =   	\otimes_{m=1}^{N-i}  M_{k_m}^{\kappa}   %
		}_{j_x=k_1,k_2,...,k_{N-i}}.
	\end{equation}	
	These assumptions are the same as presented in \cite{07FLWprotocol} to show the asymptotic optimality of converting a three-party $W$ state in a Bell pair. Here we extend them to any $\sigma_i$ state, which includes mixed states $\sigma_{i\neq0}$.

	By considering only the family of \emph{SP-LOCC transformations} $\L_i^{r,\kappa}$, we define the following lower bound
	\begin{equation}
		\LB:=\sum_i q_i(N,\epsilon) \sup_{\kappa } \sum_{j} p_{i,\Jlabel} E(\rho_{i,\Jlabel}),
		\label{eq:lb}
	\end{equation}
	which is trivially upper bounded by \eqref{eq:fig_psi} as the optimization runs over the subset of transformations. In the above, $\kappa$ is a function of $r$, since bigger and smaller protocols have different optimal sets of LOCCs.

	\subsection{SP-LOCC transformations applied to W states}
	
	Finite-state Markovian chains (FSMCs) set the mathematical framework to compute $p_{i,\Jlabel}$ and to keep track of $\rho_{i,\Jlabel}$ in a $r$-round SP-LOCC transformations. More precisely, by associating every set $\{\rho_{i,\Jlabel}^r \}_\Jlabel$ to the sampled values of a random variable $X_r$ for every $r\req0$, the process $\{X_r\}_{r\req0}$ can be interpreted as a FSMC process --- i.e., its Markovian state space $\mathcal{X}$ is a finite set, and its evolution only depends on the previous time step (see also the Appendix~\ref{sec:ap_MC} for the complete proof).
	
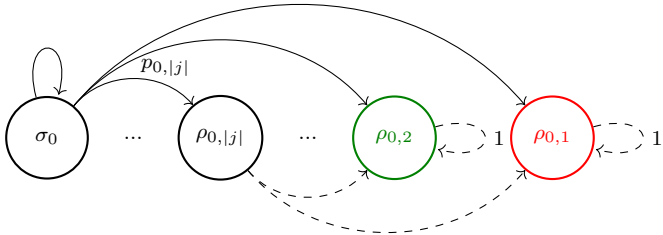
\begin{figure}
		\centering
		\usetikzlibrary{patterns}
\usetikzlibrary{automata,arrows,positioning,calc}

\newcommand{\nodedist}{2.1cm}
\newcommand{\noderadius}{.8cm}

\begin{tikzpicture}[
	% Environment Cfg
	font=\footnotesize, 
	text centered,node distance=\nodedist,
	arrow/.style ={
		-{Latex[length=3pt]},
		thick,
	},
	aMround/.style ={dashed,->}, %STYLE  for multiple rounds MC (avoid using colors)
	ms/.style ={
		circle,
		text width=\noderadius,
		draw=black,
		thick},
	edge/.style={bend left=\an, ->}
	]
	
	\def\r{.2cm};
	\def\y{5*\r};
	\def\x{.35cm}
	\def\an{50};
	
	\node[ms]    (A)                     {$\sigma_{0}$};
	\path (A) edge [loop above] node {} (A);
	
	\node[]    			(C)[right=\x of A]   {$...$}; 
	\node[ms]    (D)[right=\x of C]   {$\rho_{0,\wt{j}}$};
	\node[]    			(E)[right=\x of D]   {$...$};
	\node[ms, green]    (F)[right=\x of E]   {$\rho_{0,2}$};
	\node[ms,  red]    (G)[right of=F]   {$\rho_{0,1}$};
	
	\path (A) edge [edge] node [above,near end] {$p_{0,\wt{j}}$} (D);
	\path (A) edge [edge] node {} (F);
	\path (A) edge [edge] node {} (G);
	
%second round	

	\path (D) edge [bend left=-\an, aMround] node {} (F);
	\path (D) edge [bend left=-\an, aMround] node {} (G);

%repeat	
	\path (F) edge [loop right,aMround] node {$1$} (F);
	\path (G) edge [loop right,aMround] node {$1$} (G);

\end{tikzpicture} 
		\caption{\label{fig:MC_pure}Markovian state space and transitioning probabilities of lossless $r$-round LOCC transformations $\L_0^r$. Green (red) nodes represent Bell (separable) states.}
	\end{figure}

	The corresponding Markovian state space and transitioning probabilities of a lossless $r$-round SP-LOCC transformations $\L_0^r$ are depicted in Fig.~\ref{fig:MC_pure} with states given by
	\begin{equation}
		\rho_{0\wt{j}}= 
		\begin{cases}
			W_\wt{j} & \qif \wt{j} \req 3 \\ 
			\phi & \qif \wt{j} =2  \\ 
			\ket{0}^{\otimes N} & \qif \wt{j} = 1 ,
		\end{cases}
	\end{equation}
	where $\wt{j}$ is the number of $0$s in the string $j=k_1,k_2,...,k_{N-i}$, $W_\wt{j}$ is a $j$-qubit W state and $\phi$ is an EPR pair. This protocol is an adaptation of Fortescue and Lo's protocol \cite{07FLWprotocol} to a $N$-qubit W state (see also the Appendix~\ref{sec:ap_FL}). For the general case $i>0$ the Markovian state space increases polynomially with the number of rounds and full pictorial representation is not possible. In the following, we discuss the particular case of single-round SP-LOCC transformations. To avoid notation clutter, we will omit $r=1$ in the following sections. 
	
	\subsubsection{Single-round SP-LOCC transformations applied to W states} %$\L_i^{1,\kappa}$
	
	Given the symmetry of $\sigma_i=\Tr_{C'} W$, the form of \eqref{eq:WKraus} and the arbitrary discardment of $C'$, there will be reduced states with the same amount of entanglement. These so-named equivalent states lead to redundancies in the evaluation of \eqref{eq:lb}, the same term in the second summation is computed and optimized repeatedly, which naturally slows the computation. To eliminate these redundancies and speed up the numerical computations, as well as to obtain simpler analytical expressions, we focus on the subset of non-equivalent states and their corresponding probabilities. 
	
\begin{figure}
		\centering
		\usetikzlibrary{patterns}
\usetikzlibrary{automata,arrows,positioning,calc}

\newcommand{\nodedist}{1.6cm}
\newcommand{\noderadius}{.8cm}

\begin{tikzpicture}[
	% Environment Cfg
	font=\footnotesize, 
	text centered,node distance=\nodedist,
	arrow/.style ={
		-{Latex[length=3pt]},
		thick,
	},
	ms/.style ={
		circle,
		text width=\noderadius,
		draw=black,
		thick},
	edge/.style={bend left=\an, ->}
	]

\def\r{.2cm};
\def\y{5*\r};
\def\x{.3cm}
\def\an{50};

\node[ms]    (A)                     {$\sigma_{0}$};
\node[ms]    (B)[right of=A]   {$\rho_{0,N-1}$};
\path (A) edge [loop above] node {} (A);
\path (A) edge [edge] node {} (B);

\node[]    			(C)[right=\x of B]   {$...$}; 
\node[ms]    (D)[right=\x of C]   {$\rho_{0,\wt{j}}$};
\node[]    			(E)[right=\x of D]   {$...$};
\node[ms, green]    (F)[right=\x of E]   {$\rho_{0,2}$};
\node[ms,  red]    (G)[right of=F]   {$\rho_{0,1}$};

\path (A) edge [edge] node [above,near end] {$\wp_{0,\wt{j}}$} (D);
\path (A) edge [edge] node {} (F);
\path (A) edge [edge] node {} (G);

%%%second row
\node[ms] (sig1) [below = \y of B] {$\sigma_1$};
\node[ms] (rho1) [below right=\r and \r  of sig1] {$\rho_{1,N-1}$};

\node[]    			(C1)[right=\x of sig1]   {$...$}; 
\node[ms]    (D1)[right=\x of C1]   {$\rho_{1,\wt{j}}$};
\node[]    			(E1)[right=\x of D1]   {$...$};
\node[ms, green]    (F1)[right=\x of E1]   {$\rho_{1,2}$};
\node[ms,  red]    (G1)[right of=F1]   {$\rho_{1,1}$};

\path (sig1) edge [edge] node {} (rho1);
\path (sig1) edge [edge] node [above,near end] {$\wp_{1,\wt{j}}$} (D1);
\path (sig1) edge [edge] node {} (F1);
\path (sig1) edge [edge] node {} (G1);

\node[] (x)[below=1.5*\y of D1, rotate=90] {$...$};

%%%third row
\path node [ms,green] (D4) [below=4*\y of F] {$\sigma_{N-2}$} edge [loop above] node {$1$} (D4);

\end{tikzpicture} 
		\caption{\label{fig:MC_single}Markovian state space and transitioning probabilities of single-round LOCC transformations. Green (red) nodes represent bipartite entangled (separable) states.}
	\end{figure}
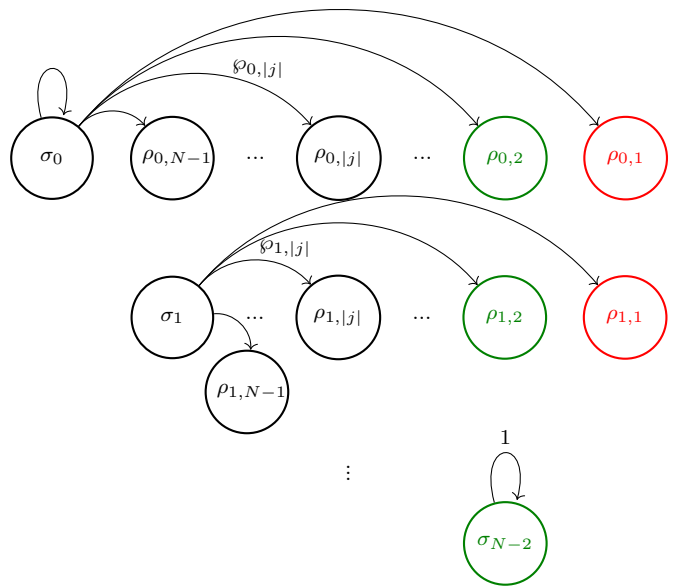

	Therefore, we define the following non-equivallent ensemble $\{\wp_{i,\wt{j}}, \rho_{i,\wt{j}} \}$, where $\wp_{i,\wt{j}}$ is the probability of obtaining any equivalent state $\rho_{i,\wt{j}}$. This restriction allows us to find the following explicit expressions
	\begin{align}
		\wp_{i,\wt{j}}&= \binom{N-i}{\wt{j}}  \frac{\wt{j}}{N-i} \kappa^{N-i-\wt{j}} (1-\kappa)^{\wt{j}-1}  \\ %j-1?
		\rho_{i,\wt{j}} & =  \binom{N-i}{\wt{j}}       \frac{\Tr_{C'} \M_{\wt{j}} \sigi \M_{\wt{j}}  }{\wp_{i,\wt{j}}}
		\label{eq:map-single}
	\end{align}
	where $\wt{j}$ is the number of $0$s in the string $j=k_1,k_2,...,k_{N-i}$. The first simplification comes from observing that permutations of $k_m$ in $j$ do not alter the number of $0$s and therefore correspond to the same state. The second is simply \eqref{eq:map_exp} expressed in terms of $\wp_{i,\wt{j}}$. These states and probabilities can be mapped to a single-trial FSMC as depicted in Fig.~ \ref{fig:MC_single}.

	\section{\label{sec:app}Application}
	
	In this section, we use the previously defined lower bound \eqref{eq:lb} to compare the performance of W and GHZ-like states in our lossy entanglement distribution scenario \ref{sec:PS}. Our results demonstrate that, in a lossy network, extracting Bell pairs from W states is more advantageous than from GHZ-like states, even though the former is only achievable probabilistically, whereas the latter can be done deterministically. Furthermore, we analytically extend these findings \ref{sec:appendix}, proving that W states serve as more effective resources in large networks.
	
	\subsection{W states' advantage in lossy networks} %main result
	
	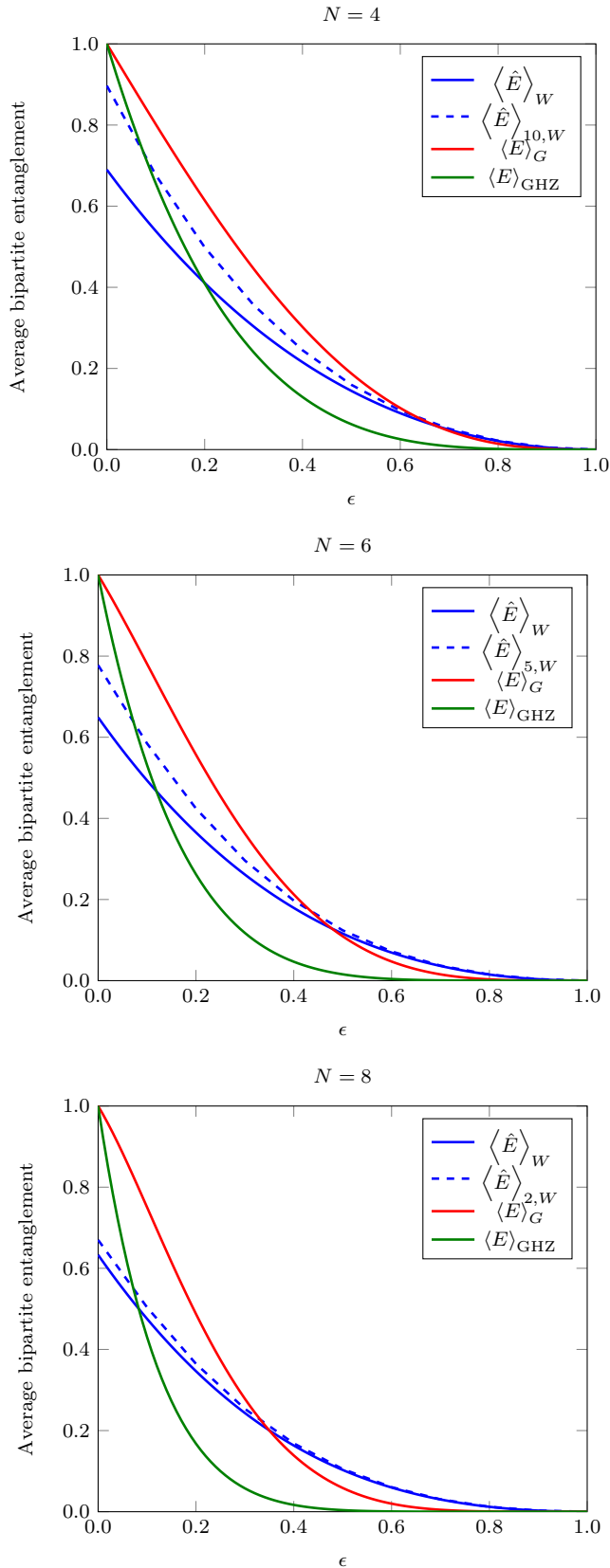
\begin{figure}%[!] %! is the only option that puts it close
		\centering
		\begin{tikzpicture}
			\node   (A)                     {\begin{tikzpicture}
	\pgfplotsset{
		tick label style={font=\footnotesize},
		label style={font=\footnotesize},
		legend style={font=\footnotesize}, 
		title style={font=\footnotesize},
%					legend image post style={black},
		every axis plot/.append style={
			line width=1pt},
		grid=minor,
		y tick label style={
			/pgf/number format/.cd,
			fixed,
			fixed zerofill,
			precision=1,
			/tikz/.cd
		},
		x tick label style={
			/pgf/number format/.cd,
			fixed,
			fixed zerofill,
			precision=1,
			/tikz/.cd
		},
		xtick={0,.2,.4,.6,.8,1},
		ytick={0,.2,.4,.6,.8,1},
		legend pos=north east
	}		
	\begin{axis}[
%		width=\imgwidth,
		%			axis y line*=right,
		%			axis x line=none,
		title={$N=4$},
		xmin=0, xmax=1,
		ymin=0, ymax=1,
		xlabel=$\epsilon$,
		ylabel=Average bipartite entanglement ,
%		legend entries={$C(W)_1$,$C(G)_1$ }% $P_W$, $P_{W\phi}$},
		legend entries={$\LBW$,$\expval{\hat{E}}_{10,W}$ ,$\EIG$,  $\EIghz$}
		]
		
		\addplot [solid,blue]  table {N4_w_r1_numeric.dat};
		\addplot [dashed,blue]  table {N4_w_10_numeric.dat};

		\addplot [solid,red]  table {N4_2-cgraph.dat};
		\addplot [solid,green]  table {N4_GHZ.dat};

	\end{axis}
	
\end{tikzpicture}};
			\node   (B) [below =.03 cm of A]   {\begin{tikzpicture}
	\pgfplotsset{
		%		every axis/.append style={line width=1.5pt},
		tick label style={font=\footnotesize},
		label style={font=\footnotesize},
		legend style={font=\footnotesize}, 
		title style={font=\footnotesize},
		%			legend image post style={black},legend style = {text=black,font=\footnotesize},
		every axis plot/.append style={
			line width=1pt},
		grid=minor,
		y tick label style={
			/pgf/number format/.cd,
			fixed,
			fixed zerofill,
			precision=1,
			/tikz/.cd
		},
		x tick label style={
			/pgf/number format/.cd,
			fixed,
			fixed zerofill,
			precision=1,
			/tikz/.cd
		},
		xtick={0,.2,.4,.6,.8,1},
		ytick={0,.2,.4,.6,.8,1},
		legend pos=north east
	}		
	\begin{axis}[
%		width=\imgwidth,
		%			axis y line*=right,
		%			axis x line=none,
		title={$N=6$},
		xmin=0, xmax=1,
		ymin=0, ymax=1,
		xlabel=$\epsilon$,
		ylabel=Average bipartite entanglement ,
%		legend entries={$\LBW$,$\EIG$,  $\EIghz$}
%		]
		legend entries={$\LBW$,$\expval{\hat{E}}_{5,W}$,
			$\EIG$,  $\EIghz$}
		]
		
		\addplot [solid,blue]  table {N6_w_r1_numeric.dat};
		\addplot [dashed,blue]  table {N6_w_4_numeric.dat};
		
		\addplot [solid,red]  table {N6_2-cgraph.dat};
		\addplot [solid,green]  table {N6_GHZ.dat};

	\end{axis}
	
\end{tikzpicture}};
			\node   (C) [below =.03 cm of B]   {\begin{tikzpicture}
	\pgfplotsset{
		%		every axis/.append style={line width=1.5pt},
		tick label style={font=\footnotesize},
		label style={font=\footnotesize},
		legend style={font=\footnotesize}, 
		title style={font=\footnotesize},
		%			legend image post style={black},legend style = {text=black,font=\footnotesize},
		every axis plot/.append style={
			line width=1pt},
		grid=minor,
		y tick label style={
			/pgf/number format/.cd,
			fixed,
			fixed zerofill,
			precision=1,
			/tikz/.cd
		},
		x tick label style={
			/pgf/number format/.cd,
			fixed,
			fixed zerofill,
			precision=1,
			/tikz/.cd
		},
		xtick={0,.2,.4,.6,.8,1},
		ytick={0,.2,.4,.6,.8,1},
		legend pos=north east
	}		
	\begin{axis}[
%		width=\imgwidth,
		%			axis y line*=right,
		%			axis x line=none,
		title={$N=8$},
		xmin=0, xmax=1,
		ymin=0, ymax=1,
		xlabel=$\epsilon$,
		ylabel=Average bipartite entanglement ,
%		legend entries={$C(W)_1$,$C(G)_1$ }% $P_W$, $P_{W\phi}$},
%		legend entries={$\LBW$,$\EIG$,  $\EIghz$}
		legend entries={$\LBW$,$\expval{\hat{E}}_{2,W}$,
		$\EIG$,  $\EIghz$}
		]
	
	\addplot [solid,blue]  table {N8_w_r1_numeric.dat};
	\addplot [dashed,blue]  table {N8_w_2_numeric.dat};
	
	\addplot [solid,red]  table {N8_2-cgraph.dat};
	\addplot [solid,green]  table {N8_GHZ.dat};

	\end{axis}
	
\end{tikzpicture}};
		\end{tikzpicture}
		\caption{\label{fig:convN}The average bipartite entanglement shared between the target parties in a lossy network, considering single-round (multiple-rounds) SP-LOCC transformations of W, GHZ and two-centered GHZ graph states as a function of the loss probability $\epsilon$ in solid (dashed) blue, red and green. From top to bottom, these quantities are depicted for a network of size $4,6$ and $8$. The intersection point common to the blue and red lines defines the \emph{lossy threshold} over which bipartite conversion of W states outperforms GHZ-like states.}
\end{figure}

We observed that probabilistically extracting Bell pairs from W states is more advantageous than doing it deterministically from GHZ-like states in lossy networks. In Fig.~\ref{fig:convN} we compare our figure of merit for W and GHZ-like states for different network sizes. It is clear that above a certain threshold in loss (when the curves intersect each other) more bipartite entanglement can be obtained on average from W than from GHZ-like states in single-round transformations, showcasing some advantage in using W states as initial resources. Moreover, such an advantage considerably increases when a higher number of rounds are allowed, as can be noticed when comparing the single- with multiple-round LOCC transformation depicted in solid and dashed lines in Fig.~\ref{fig:convN} respectively. We must stress that since we are comparing a lower bound of the average bipartite entanglement \eqref{eq:lb} extracted from the W states against the exact average bipartite entanglement extracted from GHZ-like states, the real threshold must be even lower than the one depicted in the plots. The plots also indicate that this threshold decreases as the network size increases, as observed for network sizes of $N=4,6,8$.

\subsubsection{W states' advantage in large lossy networks} %analytical result

The numerical examples of Fig.~\ref{fig:convN} indicate an inverse proportionality between the network size and the value of the threshold, as the network size increases from $4$ to $8$ the threshold decreases (from approximately $.2$ to $.1$ when comparing the W and GHZ states, for example). As we formally prove in the Appendix~\ref{sec:appendix} this trend persists for any $N$, leading to the conclusion that, for large lossy networks ($N\to\infinity$), extracting Bell pairs from W states is always more effective, according to our figure of merit, than from GHZ-like states. As detailed in the Appendix, the proof derives from the \emph{loss-robustness of W states} \cite{01WNrobust} and the inexistence of deterministic W-to-EPR state conversions~\cite{12HKlo}.

\section{\label{sec:conc}Conclusion}

In this work, we explore RED in lossy networks, focusing on whether the loss robustness of the W states outweighs the deterministic conversion of GHZ-like states into Bell pairs. We considered an $N$-partite network where entanglement is distributed through a central source over lossy channels, and network participants cooperate to establish entanglement between any two chosen parties.  To analyze this scenario, we introduced a tractable lower bound for the expected shared entanglement by optimizing our figure of merit \eqref{eq:fig_psi} over the subset of SP-LOCC transformations. By leveraging the properties of single-round SP-LOCC, we eliminated redundancies in the bound evaluation, improving general computational efficiency, and derived explicit expressions for \eqref{eq:map_exp}. 

Our results demonstrate that probabilistically extracting Bell pairs from W states is more advantageous than doing it deterministically from GHZ-like states in lossy networks. We further extended our analysis analytically, proving that W states remain more effective in large-scale networks. This has direct implications for designing optimal entanglement distribution policies --- e.g., while GHZ-like states are preferable in small, low-loss networks, W states emerge as better options in large and high-loss settings. These findings provide valuable insights into the practical deployment of lossy quantum networks, highlighting the fundamental trade-offs between probabilistic and deterministic entanglement distribution protocols.

Future endeavors include finding better bounds for the figure of merit by considereing the general optimization problem in \eqref{eq:fig_psi}, whose LOCC measurements are multi-parameterized and can be adjusted from round to round. Alternatively, diversity and multiplexing schemes \cite{junaid, Seid} could be exploited to improve RED, as entanglement distribution over a centralized QN can be framed as a multi-mode communication system \cite{compensation}.

%\onecolumn
\appendix*
\section{}
\label{sec:appendix}

In this section, we begin by motivating and presenting the definition of graph states, including the ones utilized in the paper. We then formalize the concept of resource advantage and demonstrate its use comparing W and GHZ states. Following this, we introduce the FSMC framework and provide a proof that SP-LOCC transformations can be modeled by it. Finally, we present Fortescue and Lo's entanglement distillation protocol, highlighting the similarities and differences with our own protocol.

\subsection{\label{sec:ap_graph}Graph states}

A graph is defined as a collection of vertices and a rule describing how they are connected by edges. They are often represented pictorially as points (the vertices) on a plane connected by arcs (the edges). Formally, a finite and undirected graph is defined by the pair
\begin{equation}
	G=(V,E),
	\label{eq:def-graph}
\end{equation}
where $V=\{1,...,N\}$ is the set of edges and $E\subset [V]^2$ is the set of edges and every element of $E$ is a subset of $V$ with two elements \cite{06graph_dissertation}.
In the following, we define graph states by providing physical meaning to vertices and edges --- i.e., we seek motivation for the concept of graph states in interaction patterns between quantum systems\footnote{Alternativaly, simple graphs can be associated to quantum states in terms of their stabilizer, the stabilizer formalism}. The content of this section is based on \cite{06graph_dissertation,23robgGraph}.

\subsubsection{Definition: interaction pattern}  

In the interaction pattern description, graph states are defined by providing physical meaning to vertices and edges. Specifically, vertices are associated with particles, whereas edges describe how those particles interact. For the particular case of qubits, a graph state can be regarded as a two-step procedure where qubits are prepared in some initial pure state $\ket{\psi}$ and are coupled according to the underlying interaction pattern given by the edges of $G$. Formally, for each edge $\{a,b\}\in E$, connecting qubits $a$ and $b$, a local two-particle unitary $U_{ab}=e^{- i \phi_{ab} H_{ab}}$, where $\phi_{ab}$ and $H_{ab}$ denote the coupling strength and the interaction Hamiltonian, respectively. To comply with the structure of a \emph{simple and undirected} graph $G$, these unitaries must satisfy the following constraints:
\begin{enumerate}
	\item they must commute, i.e., 
	\begin{equation}
		\bqty{ U_{ab}, U_{bc}}=0 \quad \forall a,b,c \,\in\, V;
	\end{equation}
	\item they must be symmetric, i.e.,
	\begin{equation}
		U_{ab} =  U_{ba} \quad \forall a,b \,\in\, V,
	\end{equation}
	since $G$ does not specify any ordering of the edges;
	\item they must be the same for every pair of particles, i.e., 
	\begin{equation}
		U_{ab}= U \quad \forall a,b \,\in\, V,
	\end{equation}
	since the edges are not specified with different weights.
\end{enumerate}

For qubit systems, the first condition is met by an \emph{Ising interaction pattern}. For notation convenience, we adopt the controlled phase gate
\begin{equation}
	U_{ab}(\phi_{ab}) = e^{-i\phi_{ab} H_{ab}} \quad \text{with} \quad H_{ab} := \dyad{1} \otimes\dyad{1},
\end{equation}
as done in \cite{06graph_dissertation}, which is an Ising interaction up to rotations on the $z$-axis at each qubit --- since we are interested in entanglement properties of a graph state, we can neglect and omit these rotations (see also \cite{06graph_dissertation}  for the proof). Finally, since Ising interactions are symmetric, we only need to define $\phi=\phi_{ab} \, \forall  a,b \,\in\, V$ to meet all the above constraints. As in \cite{06graph_dissertation}, we chose $\phi=\pi$ and $\ket{\psi}=\otimes_{a\in V} \ket{+}_a$ so that the resulting state $U_{ab}\ket{\psi}$ is maximally entangled (any reduced state is maximally mixed). This choice also ensures that the gate $U_{ab}$ acts on the corresponding graph creating and deleting the edge $\{a,b\}$ depending if it is contained or not in $E$. In short, we define a graph state as follows:

\begin{definition}
	Let $G=(V,E)$ be a graph. The corresponding graph state $\ket{G}$ is given by the following pure state
	\begin{equation}
		\ket{G} = \prod_{\{a,b\}\in E} U_{ab} \ket{+}^V,
		\label{eq:def-gstate}
	\end{equation}
	where 
	\begin{equation}
		U_{ab} = %
		\begin{pmatrix}
			1&0&0 & 0 \\
			0&1&0 & 0 \\
			0&0&1 & 0 \\
			0&0&0 & -1
		\end{pmatrix},
	\end{equation}
	i.e., a controlled $\sigma_z$ on qubits $a$ and $b$. 
\end{definition}
Physically, it can be pictured as a two-step preparation procedure in which the pure state $\ket{+}$ is prepared at each vertex, and a phase gate $U_{ab}$ is applied to all adjacent vertices $a,b$ in $G$. 

Next, we discuss the graph states used in this paper. 

\subsubsection{GHZ states}

The $N$-qubit GHZ state
\begin{equation}
	\ket{GHZ} = \frac{\ket{0}^{\otimes_{N}}   +     \ket{1}^{\otimes_{N}}}{\sqrt{2}}
	\label{eq:GHZ}
\end{equation}
is one of the standard examples of multiparty entangled states. As mentioned previously, these states maximally violate Bell inequalities but are sensitive to loss, and losing any qubit implies destroying all the entanglement content. %

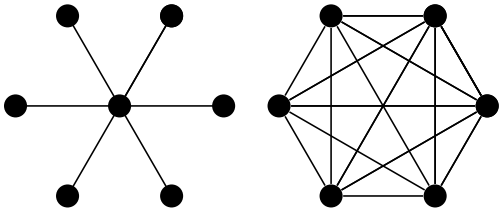
\begin{figure}
	\centering
	\begin{tikzpicture}
		\node   (A)                     {\begin{tikzpicture}[
	% Environment Cfg
	font=\footnotesize, 
	text centered,
	line/.style={semithick},
	vertex/.style={circle,fill=black,minimum size=\rnode},
	]
	\def\rnode{5}
	\def \r{.16\textwidth}
	\def \angle{360/6}
	
	\node [vertex] at (0,0) {};
		
	\draw [line] (0,0) -- (\angle*0:\r) node [vertex] {};
	\draw [line] (0,0) -- (\angle*1:\r) node [vertex] {};
	\draw [line] (0,0) -- (\angle*2:\r) node [vertex] {};
	\draw [line] (0,0) -- (\angle*3:\r) node [vertex] {};
	
	\draw [line] (0,0) -- (\angle*4:\r) node [vertex] {};
	\draw [line] (0,0) -- (\angle*5:\r) node [vertex] {};
%	\draw [line] (0,0) -- (\angle*6:\r) node [vertex] {};
	\draw [line] (0,0) -- (\angle*7:\r) node [vertex] {};

\end{tikzpicture}};
		\node   (B) [right= .2cm of A] {	
	\begin{tikzpicture}[
		% Styles
		font=\footnotesize, 
		text centered,
		line/.style={semithick},
		%		vertex/.style={circle,fill=black,minimum size=5pt, inner sep=0pt},
		vertex/.style={circle,fill=black,minimum size=\rnode},
		]
		\def\rnode{5}
		\def \r{.16\textwidth} %edge size
		\def \n{6}  % Number of nodes
		
		% Place the nodes in a circular arrangement
		\foreach \i in {0,...,7} {
			\node[vertex] (N\i) at ({360/\n * \i}:\r) {};
		}
		
		% Connect every pair of nodes to form a complete graph
		\foreach \i in {0,...,7} {
			\foreach \j in {\i,...,7} {
				\ifnum \i<\j
				\draw[line] (N\i) -- (N\j);
				\fi
			}
		}
		
	\end{tikzpicture}};
	\end{tikzpicture}
	\caption{\label{fig:graph}Graph representations of the GHZ state: the star (left) and fully connected (right) graphs.} 		
\end{figure}

The GHZ state corresponds to the star graph and the complete graph Fig.~\ref{fig:graph}. This can be seen by applying Hadamard operations to \ref{eq:GHZ} and local complementations to the star graph, which do not change the entanglement content. More specifically, by applying Hadamard operations to all qubits but one (say $a$) in \ref{eq:GHZ}, one obtains a star-graph state with central qubit $a$, which is equivalent to a complete graph up to local complementation.

As observed in \cite{23robgGraph} GHZ states are particularly sensitive (in terms of entanglement content) to loss because their corresponding star graph contains only one central vertex (root). In particular, they showed that all the correlations vanish when a root vertex is lost. This observation leads the Authors to consider redundant roots to build loss-robustness graph states as in the two-centered GHZ graph state depicted in Fig.~\ref{fig:graph_two} and detailed next.

\subsubsection{Two-centered GHZ graph states}

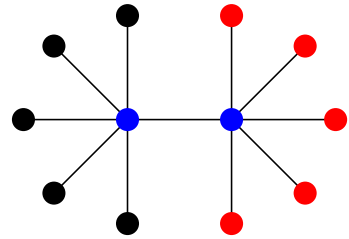
\begin{figure}
	\centering
	\begin{tikzpicture}[
	% Environment Cfg
	font=\footnotesize, 
	text centered,
	line/.style={semithick},
	vertex/.style={circle,fill=black,minimum size=\rnode},
	]
\def\rnode{5}
\def \r{.16\textwidth}
			
		\draw [line] (0,0) -- (45*2:\r) node [vertex] {};
		\draw [line] (0,0) -- (45*3:\r) node [vertex] {};
		\draw [line] (0,0) -- (45*4:\r) node [vertex] {};
		\draw [line] (0,0) -- (45*5:\r) node [vertex] {};
		\draw [line] (0,0) -- (45*6:\r) node [vertex] {};

		\draw [line] (\r,0) -- ++(45*2:\r) node [vertex,fill=red] {};
		\draw [line] (\r,0) -- ++(45*3:-\r) node [vertex,fill=red] {};
		\draw [line] (\r,0) -- ++(45*4:-\r) node [vertex,fill=red] {};
		\draw [line] (\r,0) -- ++(45*5:-\r) node [vertex,fill=red] {};
		\draw [line] (\r,0) -- ++(45*6:\r) node [vertex,fill=red] {};

		\draw [line] (0,0) node [vertex,fill=blue] {} -- ++ (\r,0) node [vertex,fill=blue] {};

\end{tikzpicture}
	\caption{\label{fig:graph_two}The loss-robustness of A two-centered GHZ graph state with $12$ qubits. The figure depicts a scenario of loss where the induced graph (obtained by removing the red vertices) is a GHZ state. If any of the two roots (blue vertices) are lost, the induced state is fully separable.} 		
\end{figure}

As defined in \cite{23robgGraph}, a two-centered GHZ graph state Fig.~\ref{fig:graph_two} is a graph state with two root vertices (the graph's centers) connected to one another, and several leaf vertices adjacent to both centers. Such states preserve their entanglement content up to losing some of the qubits, i.e., the remaining graph state still violates a Bell inequality, as carefully assessed in \cite{23robgGraph}. More precisely, it has been shown \cite{23robgGraph} that if multiple qubits adjacent to the same root are lost, the remaining state is always a GHZ state, and if qubits adjacent to both roots or the roots themselves are lost, the remaining state is (fully) separable. 

\subsection{Proof of W states' advantage over GHZ in large networks}

Here, we present two sufficient conditions so that the quantum state $\psi$ alaways outperforms the quantum state $\phi$ in an arbitrarily large lossy network, what we call \emph{$\psi$'s advantage over $\phi$}, and show that W and GHZ states satisfy them.

By exploring the monotonically decreasing behavior of \eqref{eq:fig_psi}, we propose a theorem to identify sufficient conditions in terms of $\EIpsi(\epsilon)$ and $\EIphi(\epsilon)$ so that the $N$ dimensional quantum state $\psi$ outperforms $\phi$ for any value of loss $\epsilon$ in large networks, i.e., sufficient conditions so that their nontrivial intersection\footnote{The threshold.}, $\epsilon_0$ s.t. $\EIpsi(\epsilon_0)= \EIphi(\epsilon_0)$ and $\epsilon_0\neq1$, converges to zero when $N\to \infinity$. %
The intuition behind this theorem comes from the following observations. 
\begin{itemize}
	\item If the functions $\EIpsi(\epsilon)$ and $\EIphi(\epsilon)$ have different initial values and both functions monotonically decrease to zero, there must be a non-trivial intersection $\epsilon_0<1$; and 
	\item if $\EIpsi(0) < \EIphi(0)$ and the $\EIpsi(\epsilon)$'s derivative with respect to $\epsilon$, $\dotE{\psi}:=\dexpE{\psi}$, decrease faster than $\dotE{\phi}$ for small $\epsilon$ as $N$ increases, then $\epsilon_0$ converges to zero when $N\to\infinity$.
\end{itemize} %
Assuming the first, we formally state:

\begin{theorem}[$\psi$'s advantage over $\phi$ in large lossy networks]
	Given two sequences of bounded and monotonically decreasing functions $\{\EIpsi(\epsilon)\}_N$ and $\{\EIphi(\epsilon)\}_N$ if:
	
	\begin{enumerate}
		\item $\EIpsi(0) < \EIphi(0)$ for all $N$; and
		\item $\{\dotE{\phi}\}_N$ diverges faster than $\{\dotE{\psi}\}_N$ for small $\epsilon$, e.g.,
		\begin{equation}
			\lim_{N\to \infinity} \qty[  \frac{\dotE{\psi}}{\dotE{\phi}}   ]_{\epsilon\ll1} = 0
			\label{eq:ratio-lim}
		\end{equation}
	\end{enumerate}
	then $\epsilon_0$ goes to zero when $N \to \infinity$.
\end{theorem}

Here, we prove W's advantage over GHZ states, i.e., we show that W and GHZ states, as $\psi$ and $\phi$ respectively, satisfy the above theorem.

\begin{proof} 
	Following the above theorem, we need to show that:
	\begin{enumerate}
		\item $	\EIw(0)  < \EIghz(0)$ for all $N$; and
		\item $\{\dotE{\ghz}\}_N$ diverges faster than $\{\dotEW\}_N$ for small $\epsilon$, e.g.,
		\begin{equation}
			%			\lim_{N\to \infinity} \qty[  \frac{\dotE{\phi}}{\dotE{\psi}}   ]_{\epsilon=0} = \infinity
			\lim_{N\to \infinity} \qty[  \frac{\dotEW}{\dotE{\ghz}}   ]_{\epsilon=0} = 0
			\label{eq:ratio-lim}
		\end{equation}
	\end{enumerate}

	We first show the first condition is met. From \eqref{eq:q_i} we have that $q_0(0,N)=1 $ and $q_i(0,N)=0 \, \forall i>0$, therefore:
	\begin{equation}
		\EIw(0) = \barEum{\sigma_0} < 1  
	\end{equation}
	and 
	\begin{equation}
		\EIghz(0) = \barEum{\ghz} = 1,
	\end{equation}
	which follows respectively from the fact that W (GHZ) states are probabilistically (deterministically) transformed in a Bell pair \cite{12HKlo}. Combining the above equations we find
	\begin{equation}
		\EIw(0)  < \EIghz(0),
	\end{equation}
	as we wanted to prove.

	The second condition follows from the loss-robustness of W states \cite{01WNrobust}, i.e.,
	\begin{equation}
		\lim_{N\to\infinity} F(\sigma_0^{N-1},\sigma_1^N )=0,
		\label{eq:LR_w}
	\end{equation}
	where $F$ is the fidelity. Since only $q_i$ depends on $\epsilon$, we have
	\begin{align}
		\EIxprimeW(0)&= q_0'(N,0)\barEum{\sigma_0^N} + q_1'(N,0)\barEum{\sigma_1^N}  \\
									&=-(\MM)\qty[  \barEum{\sigma_0^N}- \barEum{\sigma_1^N}  ] \nonumber\\
									&=-(\MM)\qty[  \sup_{ \L_i^1 } \wp_{0N} E(\sigma_0^N) + \barEum{\sigma_0^{N-1}}     - \barEum{\sigma_1^N}  ],
		\label{eq:w}
	\end{align}
	where we have used \eqref{eq:fig_sig} to expand the first term in the second line. Similarly,
	\begin{align}
		\EIxprime{\ghz}(0) &= -(\MM)
		\label{eq:ghz}
	\end{align}
	since $\barEum{\sigma_0^N}=1$ and $\barEum{\sigma_i^N}=0 \, \forall i>0$ in this case. The ratio $ \dotEW/\dotE{\ghz}$ simplifies to
	\begin{equation}
		\sup_{ \L_i^1 } \wp_{0N} E(\sigma_0^N) + \barEum{\sigma_0^{N-1}}     - \barEum{\sigma_1^N}.
	\end{equation}
	which goes to zero when $N\to \infinity$ since $\wp_{0N}$ decreases with $N$ (check \eqref{eq:map-single}) while $E(\sigma_0^N) \in [0,1]$, and $\sigma_0^{N-1} \to \sigma_1^{N}$ when $N\to \infinity$ as assumed in \eqref{eq:LR_w}.
	
\end{proof}

\subsection{\label{sec:ap_MC}Discrete-time finite-state Markov chains}

A Markov chain (MC) is a stochastic process defined at integer values of time $r=0,1,2,...$, that is, for every $r\req0$, there is a random variable $X_r$, the chain state at time $r$ (see also \cite{book:markov}). 

\begin{definition}[Markov chain process]
	The evolution of a MC is defined by $ \{X_r \}_{r\req 0}$, where:
	\begin{enumerate}
		\item the collection of all possible values of all the $X_r$, the Markovian state space $\mathcal{X}$, is a countable set;
		\item the sampled values of each $X_r$ depends only on the most recent (chain) state $X_{r-1}$. More specifically, for all positive $r$, 
		\begin{equation}
			P[X_r \mid X_{r-1},  X_{r-2},...,X_0] = P[X_r \mid X_{r-1}],
		\end{equation}
		where the initial (chain) state $X_0$ has an arbitrary distribution.
	\end{enumerate}
\end{definition}

In such an MC process it is often useful to compute the probability of going to state $j$ in $r$ steps starting in the state $i$, i.e., $P[X_r=j \mid X_{0} = i]$. From the Chapman-Kolmogorov equation, we have that 
\begin{equation}
	P[X_r=j \mid X_0 = i] = (P^r)_{ij},
	\label{eq:CKeq}
\end{equation}
where $P$ is the transition probability matrix, whose elements are $P_{ij}=P[X_1 = j \mid X_0 =i]$. That is, $P[X_r=j \mid X_{0} = i]$ equals the $i,j$ element of the $r$th power of matrix $P$. A finite-state MC is a MC whose Markovian state space is finite.

FSMCs set the mathematical framework to compute $p_{i,\Jlabel}$ and to keep track of $\rho_{i,\Jlabel}$ in a $r$-round LOCC transformations $\L_i^r$. More precisely, by associating every set $\{\rho_{i,\Jlabel}^r \}_\Jlabel$ to the sampled values of a random variable $X_r$ for every $r\req0$, the process $\{X_r\}_{r\req0}$ can be interpreted as an FSMC process --- that is, its Markovian state space $\mathcal{X}$ is a finite set, and its evolution depends only on the previous time step.

\begin{proof}
	The first condition follows directly from the definition \eqref{eq:rho}, i.e., $\{\rho_{i,\Jlabel}^r\}$ corresponds to a finite set of LOCC operations, indexed by $\Jlabel$. The second condition is satisfied by definition, i.e., for all $\rho_{i,\mathbf{m}}$ and $\rho_{i,\mathbf{l}} \in \mathcal{X}$ we define the probability of going to state $\rho_{i,\mathbf{m}}$, starting from $\rho_{i,\mathbf{l}}$, as 
	\begin{equation}
		P[X_r = \rho_{i,\mathbf{m}} \mid X_{r-1} = \rho_{i,\mathbf{l}}] := \Tr \mathcal{M}_\mathbf{m}^1 \mathcal{M}_\mathbf{m}^1 \rho_{i,\mathbf{l}},
		\label{eq:tran_prob}
	\end{equation}
	where $\mathcal{M}_\mathbf{m}^1$ is a single-round global measurement given by \eqref{eq:rounds}. In other words, \eqref{eq:tran_prob} corresponds to the probabilities of a single-round LOCC transformation \eqref{eq:map} acting on $\rho_{il}$.
\end{proof}

With this definition, 
\begin{equation}
	p_{i,\Jlabel}^r \equiv P[X_r = \rho_{0\Jlabel} \mid X_0 = \sigma_i] = (P^r)_{i\Jlabel},
\end{equation}
where the last equality follows from \eqref{eq:CKeq}.

\subsection{\label{sec:ap_FL}Fortescue and Lo's protocol}

In \cite{07FLWprotocol} Fortescue and Lo present a probabilistic entanglement distillation protocol able to distill Bell pairs from a three-qubit W state.  In this protocol, the three parties measure their share of the entangled state, using the set of Kraus operators defined in \eqref{eq:WKraus}, if:
\begin{enumerate}
	\item all parties get outcome $0$, they will share the same W state and repeat the protocol;
	\item two of the three parties get outcome $0$, they will share a Bell pair and successfully terminate the protocol;
	\item only one the three parties get outcome $0$, they will share a separable state $\ket{0}\ket{0}\ket{0}$ and unsuccessfully terminate the protocol.
\end{enumerate} 
These events correspond to repeat, success, and failure events and occur with probability $(1-\kappa)^2$, $2\kappa^2(1-\kappa^2)$ and $\kappa^2$. They found that after $r$ executions of the protocol, the maximum average entanglement shared among an unspecified pair of parties is $\frac{r}{1+r}$. 

Our protocol is very similar. All parties measure their systems with the same set of Kraus operators \eqref{eq:WKraus}, but success is deemed not only to Bell states but to any final entangled state, which includes the bipartitions of W states. Moreover, we also apply them to lossy $N$-party W states $\sigma_{i\neq0}$.

\bibliography{ref_v2.bib}
\end{document}